\documentclass{elsart}

\newcounter{bla}

\newcommand{\bea}{\begin{eqnarray}}
\newcommand{\eea}{\end{eqnarray}\noindent}
\newcommand{\nn}{\nonumber}
\newcommand{\calst}{\mbox{$\cal S$}}

\def\eps{\epsilon}
\usepackage{amssymb,amsmath}
\usepackage{epsfig}

\begin{document}

\bibliographystyle{JHEP}

\begin{frontmatter}

\hfill{Edinburgh 2011/02}\\
\hfill{LAPTH-xy/11}\\
\hfill{IPPP/11/04, DCPT/11/08}\\ 
\hfill{Nikhef/2011-002}\\
\hfill{MPP-2011-5}

\title{Golem95C: A library for one-loop integrals with complex masses}

\author[a]{G.~Cullen},
\author[b]{J.-Ph.~Guillet},
\author[c]{G.~Heinrich},
\author[d]{T.~Kleinschmidt},
\author[b]{E.~Pilon},
\author[e]{T.~Reiter},
\author[d]{M.~Rodgers}


\address[a]{School of Physics, The University of Edinburgh, Edinburgh EH9 3JZ, UK}
\address[b]{LAPTH, Universit\'e de Savoie and CNRS, Annecy-le-Vieux, France}
\address[c]{Max-Planck-Institut f\"ur Physik, F\"ohringer Ring 6, 80805 M\"unchen, Germany}
\address[d]{Institute for Particle Physics Phenomenology,
        University of Durham, \\Durham, DH1 3LE, UK}
\address[e]{Nikhef, Science Park 105, 1098 XG Amsterdam, The Netherlands}

\begin{abstract}
We present a program for the numerical evaluation of scalar 
integrals and tensor form factors 
entering the calculation of one-loop amplitudes 
which supports the use of complex masses in the loop integrals.
The program is built on an earlier version of the 
golem95 library, which performs the reduction 
to a certain set of basis integrals
using a formalism where inverse Gram determinants can be avoided.
It can be used to calculate one-loop amplitudes
with arbitrary masses in an algebraic approach as well as in the context of 
unitarity-inspired numerical reconstruction of the integrand.
\begin{flushleft}
\end{flushleft}


\end{abstract}

\end{frontmatter}

{\bf NEW VERSION PROGRAM SUMMARY}

\begin{small}
\noindent
{\em Manuscript Title: Golem95C: A library for one-loop integrals with complex masses}                                       \\
{\em Authors: G.~Cullen, J.-Ph.~Guillet,  G.~Heinrich, T.~Kleinschmidt, E.~Pilon, T.~Reiter, M.~Rodgers}                                                \\
{\em Program Title: golem95-1.2.0}                                          \\
{\em Licensing provisions:}      none                             \\
{\em Programming language:}     Fortran95                              \\
{\em Computer: Any computer with a  Fortran95 compiler }   \\
{\em Operating system:}      Linux, Unix                                   \\
{\em RAM:} RAM used per integral/form factor  is insignificant                                             
{\em Classification:} 4.4, 11.1                                       \\
{\em External routines/libraries:}   some finite scalar integrals are called from OneLOop\,\cite{vanHameren:2009dr,vanHameren:2010cp}, 
 the option to call them from LoopTools\,\cite{Hahn:1998yk,Hahn:2010zi} is also implemented.                               \\
{\em Catalogue identifier of previous version:}   AEEO\_v1\_0           \\
{\em Journal reference of previous version:}    Comput. Phys. Commun. 180 (2009) 2317.   \\
{\em Does the new version supersede the previous version?:} yes   \\
{\em Nature of problem:} Evaluation of 
one-loop multi-leg integrals occurring in the calculation of next-to-leading order corrections
to scattering amplitudes in elementary particle physics. In the presence of 
massive particles in the loop, propagators going on-shell can cause singularities which should be regulated
to allow for a successful evaluation.
\\
{\em Solution method:} Complex masses can be used  in the loop integrals to 
stand for a width of an unstable particle,  regulating the
singularities by moving the poles away from the real axis. \\
   {\em Reasons for the new version:} The previous version was restricted to massless 
particles in the loop.\\
   {\em Summary of revisions:} Real and complex masses are supported, 
a general $\mu$ parameter for the renormalisation scale is introduced, 
improvements in the caching system and the user interface.\\
{\em Running time:} Depends on the nature of the problem. 
A single call to a rank 6 six-point form factor 
at a randomly chosen kinematic point, using complex masses,  takes 0.06 seconds on an 
Intel Core 2 Q9450 2.66\,GHz processor. \\

\end{small}

\newpage

\hspace{1pc}
{\bf LONG WRITE-UP}

\section{Introduction}
Collider experiments at the TeV scale, in particular
the LHC experiments,  are 
expected to shed light on the mechanism of electroweak symmetry breaking 
and possibly guide us to  a more complete theory of 
fundamental interactions than the present Standard Model.
In order to achieve these goals, predictions for signal as well as background 
processes should be well under control, necessitating calculations at
next-to-leading order (NLO) accuracy or beyond. 

Over the last few years, enormous progress has been made to 
push the calculation of NLO corrections towards a higher number of 
particles in the final states, i.e. to ``multi-leg" amplitudes, 
in QCD as well as in the electroweak sector. 
For  reviews see e.g. \cite{Bern:2008ef,Binoth:2010ra}.

Nowadays, the efforts  are also increasingly being focused on the goals of {\em automating} 
multi-leg one-loop calculations and making them {\em publicly available}. 
Recent public programs with emphasis on  multi-leg one-loop calculations
are e.g. 
CutTools\,\cite{Ossola:2007ax},
Samurai\,\cite{Mastrolia:2010nb,Heinrich:2010ax},
other public programs which have been optimized for less than four particles in the final state 
are e.g. 
FeynArts/FormCalc\,\cite{Hahn:1998yk,Hahn:2000kx,Hahn:2006qw,Hahn:2010zi}, 
MCFM\,\cite{Campbell:2002tg}, VBFNLO\,\cite{Arnold:2008rz}.

An important ingredient for such programs is an integral library containing the one-loop 
integrals which are the basic building blocks of  any one-loop amplitude unless it is
calculated purely numerically. 
Several libraries are available to date: FF\,\cite{vanOldenborgh:1989wn,vanOldenborgh:1990yc}, Looptools\,\cite{Hahn:2006qw}, 
QCDLoop\,\cite{Ellis:2007qk}, 
OneLOop\,\cite{vanHameren:2010cp},  
golem95\,\cite{Binoth:2008uq}, Hexagon.F\,\cite{Diakonidis:2010rs}.
A code for the calculation of one-loop four-point functions with complex masses ({\tt D0C}) 
can be found in \cite{Nhung:2009pm}.  The latter has been integrated into the LoopTools 
library\,\cite{Hahn:2010zi} where the complex version of  infrared finite 
integrals with less than four legs are already implemented. 
A complete set of scalar four-point integrals, both in dimensional and in mass regularisation 
and valid also for complex masses 
can be found in \cite{Denner:2010tr} in analytic form. 

The calculation of scalar one-loop integrals has a long tradition of pioneering work, see e.g. 
\cite{tHooft:1978xw,Fabricius:1979tb,vanOldenborgh:1989wn,Beenakker:1988jr,Denner:1991qq,Bern:1992em,Bern:1993kr}.  
For processes involving unstable particles, these integrals are also required for 
complex internal masses, in order to be able to work within the so-called 
``complex-mass scheme" developed in Refs.\,\cite{Denner:1999gp,Denner:2005fg}.
For calculations beyond one loop, complex values for invariants derived from external momenta are 
also required\,\cite{Actis:2008uh,Passarino:2010qk}, 
but we will concentrate on one-loop corrections here.

In this article, we present an extension  of the tensor and scalar library 
of Ref.\,\cite{Binoth:2008uq} to integrals with 
arbitrary masses, in particular also complex masses.
Furthermore, we extend the approach which was previously based solely on form factors 
as building blocks of the amplitude to an approach which is 
useful in the context of reconstruction of the integrand using  $D$-dimensional unitarity. 

This article is organized as follows. In Section 2, we review the theoretical background, 
with particular emphasis on the treatment of potential numerical instabilities. 
In subsection 2.4 we give an example to demonstrate how the introduction of complex masses 
can cure Landau singularities stemming from on-shell massive particles in the loop.
Subsection 2.5 is dedicated to the new feature of {\tt golem95} to be used in the context of 
a numerical reconstruction of amplitudes at the integrand level.
Section 3 gives a brief overview of the software structure, while a detailed 
description of the individual  software components is provided in Section 4.
The installation instructions can be found in Section 5, followed by a listing of the 
examples which are new in this version in Section 6, before we conclude.

\vspace*{3mm}

\section{Theoretical background}

The program is an update of the tensor and scalar integral 
library described in more detail in  Ref.~\cite{Binoth:2008uq}, 
based on the formalism  developed in Refs.~\cite{Binoth:2005ff,Binoth:1999sp}
to reduce tensor integrals to a convenient set of basis integrals. 
Similar reduction schemes can be found e.g. in 
Refs.~\cite{Bern:1992em,Duplancic:2003tv,Giele:2004iy,delAguila:2004nf,vanHameren:2005ed,Denner:2005nn,Fleischer:2010sq}.
Here we will describe the theoretical framework only  briefly 
and focus on the new features of the program.

\subsection{Form Factors}

Tensor integrals can be divided into a part containing the Lorentz structure 
and a part consisting of scalar quantities, which we call {\it form factors}.

We define an $N$-point tensor integral of rank $r$ in $D=4-2\eps$ 
dimensions as 
\begin{eqnarray}
I^{D,\,\mu_1\ldots\mu_r}_N(a_1,\ldots,a_r) = 
\int \frac{d^D q}{i \, \pi^{D/2}}
\; \frac{q_{a_1}^{\mu_1}\,\dots  q_{a_r}^{\mu_r}}{
(q_1^2-m_1^2+i\delta)\dots (q_N^2-m_N^2+i\delta)}
\label{eq0}
\end{eqnarray} 
where $q_a=q+r_a$, $q$ is the loop momentum, and 
$r_a$ is a combination of external momenta.
Using the shift invariant vectors 
\begin{equation}
\Delta_{ij}^\mu=
r_i^\mu - r_j^\mu\;,
\end{equation}
we can write

\begin{multline}\label{eq:formfactordef}
I^{D,\mu_1\ldots\mu_r}_N(a_1,\ldots, a_r; S)=
\sum_{j_1,\ldots,j_r\in S}
   \left[\Delta_{j_1\cdot}^{\cdot}\cdots\Delta_{j_r\cdot}^{\cdot}%
   \right]^{\{\mu_1\ldots\mu_r\}}_{\{a_1\ldots a_r\}}
   A^{N,r}_{j_1\ldots j_r}(S)
\\
+ \sum_{j_1,\ldots,j_{r-2}\in S}
   \left[g^{\cdot\cdot}%
   \Delta_{j_1\cdot}^{\cdot}\cdots\Delta_{j_{r-2}\cdot}^{\cdot}%
   \right]^{\{\mu_1\ldots\mu_r\}}_{\{a_1\ldots a_r\}}
   B^{N,r}_{j_1\ldots j_{r-2}}(S)
\\
+ \sum_{j_1,\ldots,j_{r-4}\in S}
   \left[g^{\cdot\cdot}g^{\cdot\cdot}%
   \Delta_{j_1\cdot}^{\cdot}\cdots\Delta_{j_{r-4}\cdot}^{\cdot}%
   \right]^{\{\mu_1\ldots\mu_r\}}_{\{a_1\ldots a_r\}}
   C^{N,r}_{j_1\ldots j_{r-4}}(S)\;.
\end{multline}

The notation $[\cdots]^{\{\mu_1\cdots\mu_r\}}_{\{a_1\cdots a_r\}}$ 
stands for the distribution of the $r$ Lorentz indices $\mu_i$, and the momentum 
labels $a_i$ to the vectors $\Delta_{j\,a_i}^{\mu_i}$ and metric tensors 
in all distinguishable ways. 
$S$ denotes an ordered 
set of propagator labels, corresponding to the momenta forming 
the kinematic matrix ${\cal S}$, defined by 
\bea
\calst_{ij} &=&  (r_i-r_j)^2-m_i^2-m_j^2\;\quad ; \;\quad i,j\in\{1,\ldots,N\}\;.
\label{eqDEFS}
\eea
The form factors are linear combinations of so-called 
reduction coefficients derived from the matrix ${\cal S}$ and $N$-point integrals
with $N\leq 4$.
The kinematic matrix ${\cal S}$ is related to the Gram matrix $G_{ij}$ 
($i,j=1,\ldots ,N-1$ for $r_N=0$)
by 
\bea
\det G&=& (-1)^{N+1}B\,\det{\cal S}\;,\;B=\sum_{i,j=1}^N {\cal S}^{-1}_{ij}\label{Bdef}
\;.
\eea

\subsection{Integrals}

The {\tt golem95} program uses the fact that tensor integrals 
are related to  Feynman parameter integrals with Feynman parameters in the numerator.
A scalar integral, after Feynman parametrisation, can be written as 
\bea
I^D_N(S) &=& (-1)^N\Gamma(N-\frac{D}{2})\int \prod_{i=1}^N dz_i\,
\delta(1-\sum_{l=1}^N z_l)\,\left(R^2\right)^{\frac{D}{2}-N}\nn\\
&& R^2 =  
-\frac{1}{2} \sum\limits_{i,j=1}^N z_i\,\calst_{ij}  z_j\,\,-i\delta
\;.
\label{isca2}
\eea

The general relation between tensor integrals and parameter integrals 
with Feynman parameters in the numerator is
 well known~\cite{Davydychev:1991va,Bern:1992em,Binoth:1999sp}
\bea
&&I^{D,\,\mu_1\ldots\mu_r}_N(a_1,\ldots,a_r\,;S) 
 =  
(-1)^r \sum_{m=0}^{[r/2]} \left( -\frac{1}{2} \right)^m\nn\\ 
&&
\sum_{j_1\cdots j_{r-2m}=1}^N \left[ 
 (g^{..})^{\otimes m}\,\Delta_{j_1\cdot}^{\cdot} \cdots \Delta_{j_r\cdot}^{\cdot}
\right]^{\{\mu_1\cdots\mu_r\}}_{\{a_1\cdots a_r\}}
\;
 I_N^{D+2m}(j_1 \ldots ,j_{r-2m}\,;S)\;,
\label{eq32}
\eea 
where $I_N^{D+2m}(j_1 \ldots ,j_{r-2m}\,;S)$ 
 is an integral with Feynman parameters in the numerator.
 $[r/2]$ stands for the nearest integer less or equal to $r/2$ and the symbol 
 $\otimes m$ indicates
 that $m$ powers of the metric tensor are present.
Feynman parameter integrals corresponding to 
diagrams where propagators $j_1,\dots,j_m$ are omitted, or {\it pinched},
with respect to the ``maximal" topology 
can be defined as
\bea
&&I^D_N(j_1,\dots,j_r;S\setminus \{l_1,\dots,l_m\}) =(-1)^N\Gamma(N-\frac{D}{2})
\nn\\
&& 
\int \prod_{i=1}^N dz_i\,
\delta(1-\sum_{k=1}^N z_k)\,
\delta(z_{l_1})\dots \delta(z_{l_m})z_{j_1}\dots z_{j_r}\left(R^2\right)^{D/2-N}\;.
\label{isca_pinch}
\eea


The program {\tt golem95} reduces the integrals internally to a set of 
basis integrals, i.e. the endpoints of the reduction 
(they do not form a basis in the mathematical sense, 
as some of them are linearly dependent).
The choice of the basis integrals can have important effects on the numerical stability 
in certain kinematic regions, as will be explained below.
Our reduction endpoints are 
4-point functions in 6 dimensions
$I_4^6$, which are IR and UV finite, UV divergent 4-point functions in
$D+4$ dimensions, and various 2-point and 3-point functions, some of
the latter  with Feynman parameters in the numerator. This provides us with a 
convenient  separation of IR and UV divergences, as the IR poles are 
exclusively contained in
the triangle functions. Explicitly, our reduction 
basis is given by integrals of the type 
\begin{eqnarray}\label{basisintegral}
I^{D}_3(j_1, \ldots ,j_r) &=& 
-\Gamma \left(3-\frac{D}{2} \right) \, \int_{0}^{1} 
\prod_{i=1}^{3} \, d z_i \, \delta(1-\sum_{l=1}^{3} z_l) 
\, \frac{z_{j_1} \ldots z_{j_r}}{ 
(-\frac{1}{2}\, z \cdot \calst
\cdot z-i\delta)^{3-D/2}}\;,\nn\\
I^{D+2}_3(j_1) &=& 
-\Gamma \left(2-\frac{D}{2} \right) \, \int_{0}^{1} 
\prod_{i=1}^{3} \, d z_i \, \delta(1-\sum_{l=1}^{3} z_l) 
\, \frac{z_{j_1}}{ 
(-\frac{1}{2}\, z \cdot \calst
\cdot z-i\delta)^{2-D/2}}\;,\nn\\
I^{D+2}_4(j_1, \ldots ,j_r) &=& 
\Gamma \left(3-\frac{D}{2} \right) \, \int_{0}^{1} 
\prod_{i=1}^{4} \, d z_i \, \delta(1-\sum_{l=1}^{4} z_l) 
\, \frac{z_{j_1} \ldots z_{j_r}}{ 
(-\frac{1}{2}\, z \cdot \calst
\cdot z-i\delta)^{3-D/2}}\;,\nn\\
I^{D+4}_4(j_1) &=& 
\Gamma \left(2-\frac{D}{2} \right) \, \int_{0}^{1} 
\prod_{i=1}^{4} \, d z_i \, \delta(1-\sum_{l=1}^{4} z_l) 
\, \frac{z_{j_1}}{ 
(-\frac{1}{2}\, z \cdot \calst
\cdot z-i\delta)^{2-D/2}}\;,\nn\\
\end{eqnarray}
where $r^{\rm{max}}=3$, as well as $I^{D}_3,I^{D+2}_3,I^{D+2}_4,I^{D+4}_4$ 
with no Feynman parameters in the numerator, 
and two-point and one-point functions.

Note that $I^{D+2}_3$ and $I^{D+4}_4$ are UV divergent, while  $I^{D}_3$
can be IR divergent. In the code, the integrals are represented as 
arrays containing the coefficients of their Laurent expansion in $\epsilon=(4-D)/2$.

We would like to emphasize that the program also can be used as a library for 
scalar master integrals.

\subsection{Treatment of potential numerical instabilities}

\subsubsection{Spurious singularities due to inverse Gram determinants}

Further reduction of the integrals in eqs.~(\ref{basisintegral}) to integrals 
with 
no Feynman parameters in the numerator  introduces factors of $1/B$, i.e. 
inverse Gram determinants. A particular feature of {\tt golem95} is the fact that 
the above integrals are {\it not} reduced to scalar basis integrals
in cases where $B$ becomes small, thus avoiding problems with small inverse determinants.
In these cases, the above integrals are evaluated numerically. 
As $B= (-1)^{N+1}\det(G)/\det({\cal S})$ is a dimensionful quantity, 
the switch to the numerical evaluation of the basis integrals is 
implemented such that the value of the dimensionless parameter $\hat{B}$ is tested, where 
\begin{equation}
\hat{B}=B\times (\rm{largest \; entry \; of \;} {\cal S})\;.
\label{bhat}
\end{equation}
If $\hat{B}>\hat{B}^{\rm{cut}}$, the reduction is performed, else the program 
switches to the direct numerical evaluation of the integral.
The default value is $\hat{B}^{\rm{cut}}=0.005$.
In particular, we use a certain one-dimensional parameter representation
here, obtained after performing two integrations analytically.
In this way one can use deterministic integration routines, 
leading to a fast and precise numerical evaluation. 
The relative error to be achieved in the numerical integration has been set to 
the default value $10^{-8}$. If this precision has not been reached, the program will 
write a message to the file {\tt error.txt}. 

This switch to a direct numerical evaluation will be done automatically for all triangle integrals, and 
for box integrals with massless propagators and up to three off-shell legs.
For box diagrams with massive internal propagators the one-dimensional parameter representation
is not yet implemented, but will be provided in a forthcoming version.

In some cases, calculating in double precision Fortran may not be sufficient.
The code is designed such that  it can be compiled in quadruple precision as well.

\subsubsection{Landau Singularities}

After Feynman parametrisation and momentum integration, the denominator ${\cal D}$ of any one-loop 
integral is given by ${\cal D}=-z\cdot {\cal S}\cdot z/2-i\delta$
(see eq.\,(\ref{isca2})).  
Necessary conditions for a Landau singularity to occur in a one-loop integral 
thus can be expressed as 
\begin{equation}
\det {\cal S}=0\;,\; z_i \geq 0 \;,\;   z\cdot {\cal S}\cdot z<0  \;.
\end{equation}
The {\it leading Landau singularity}, 
corresponding to $\det {\cal S}=0$ and $z_i>0$ for all $i$,
occurs if all particles in the loop go simultaneously on-shell. 
Sub-leading Landau singularities occur if a sub-matrix of ${\cal S}$ has a vanishing determinant
and at least one of the Feynman parameters $z_i$ is zero, corresponding to pinched diagrams.
Leading Landau singularities are also called {\it anomalous thresholds} in the 
literature\,\cite{BjorkenDrell,Goria:2008ny},  this term stemming from the fact that 
they only occur in the physically allowed phase space region if a number of rather special 
kinematical conditions   are fulfilled. An example will be given below.

\subsection{Complex masses}

\begin{figure}[htb]
\unitlength=1mm
\begin{picture}(120,50)
\put(25,5){\includegraphics[width=8.5cm]{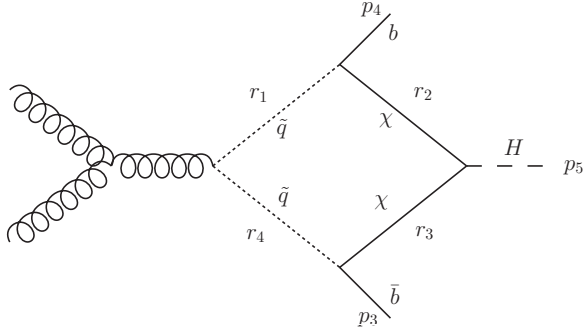}}
\end{picture}
\caption{A diagram where all propagators can go simultaneously on-shell to
develop a leading Landau singularity which can be regulated by introducing complex masses.}
\label{fig:Hbb}
\end{figure}

In order to demonstrate how the introduction of complex masses, 
standing for  a width in the case of unstable particles in the loop, 
regulates the  Landau singularities, 
we choose the example of a box diagram contributing to the production of a heavy neutral 
Higgs boson and a $b\bar{b}$ pair in gluon fusion 
in the context of  supersymmetry, where the loop contains two 
squarks (sbottoms) and two neutralinos, as shown in Fig.\,\ref{fig:Hbb}.

Denoting the momentum of the Higgs boson by $p_5$ and the momenta of the 
two b-quarks by $p_3$ and $p_4$, the kinematic matrix ${\cal S}^{(4)}$ associated with the 
four-point diagram
 is symmetric under  $s_{35}\leftrightarrow s_{45}$ ($s_{ij}=(p_i+p_j)^2)$. 
To make the singularity structure apparent, we will scan the 
different thresholds and singularities as a function of the 
invariant  $s_{45}$. The $b$-quarks are taken to be massless in this example.
Similar investigations have been worked out in detail in \cite{Denner:1996ug,Boudjema:2008zn}.

\begin{figure}[htb]
\unitlength=1mm
\begin{center}
\begin{picture}(150,100)
\put(5,90){\includegraphics[width=8.5cm,angle=-90]{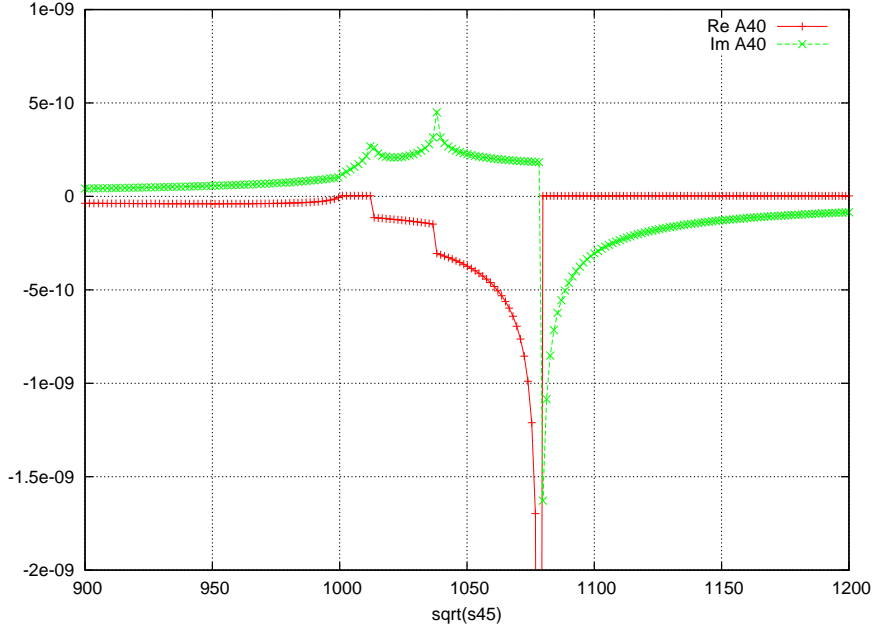}}
\end{picture}
\end{center}
\caption{Singularity structure of the scalar four-point function $A^{4,0}$ (real masses) contained in the diagram 
of Fig.\,\ref{fig:Hbb} for $m_H=450$\,GeV, $m_{\tilde{q}}=800$\,GeV, $m_\chi=200$\,GeV, $\sqrt{s}=1700$\,GeV, $900\,{\rm{GeV}}\leq s_{45}\leq 1200$\,GeV.}
\label{fig:Hbbsing}
\end{figure}

Solving the phase space constraint $\det G\geq 0$ \; ($G_{ij}=2p_ip_j, i=3,4,5$) 
for $s_{45}$ leads to the following boundaries for the physically allowed phase space:
\bea
M_H^2\,\frac{s_{12}}{s_{35}}&\leq& s_{45} \leq M_H^2+s_{12}-s_{35}\label{psboundaries}\\
M_H^2&\leq& s_{35} \leq s_{12}\nn\;.
\eea
The leading Landau singularity is characterised  by all particles in the loop 
going simultaneously on-shell, leading to $\det{\cal S}^{(4)}=0$. 
The determinant of the kinematic matrix ${\cal S}^{(4)}_{ij}$
can be written as \cite{Boudjema:2008zn}
\bea
\det {{\cal S}}^{(4)}&=&\lambda(s_{35},m_{\tilde{q}}^2,M_\chi^2)\,(s_{45}-s_{45}^0)^2+
\det {{\cal S}}^{(3)}_{\hat{r}_2}\,\det {{\cal S}}^{(3)}_{\hat{r}_4}\label{detQ4}\\
\lambda(x,y,z)&=&x^2+y^2+z^2-2\,(xy+xz+yz)\;.\nn
\eea
The determinants $\det {{\cal S}}^{(3)}_{\hat{r}_2}$ and $\det {{\cal S}}^{(3)}_{\hat{r}_4}$
correspond to diagrams where the propagators $r_2$ and $r_4$ respectively in Fig.\,\ref{fig:Hbb} 
are pinched. The Kaellen function 
$\lambda(s_{35},m_{\tilde{q}}^2,m_\chi^2)$ is (minus) the determinant of the kinematic matrix 
associated with a two-point function where both $r_1$ and $r_3$ are pinched.
$s_{45}^0$ is the solution of the equation $\det {{\cal S}}^{(4)}-\det {{\cal S}}^{(3)}_{\hat{r}_2}\,\det {{\cal S}}^{(3)}_{\hat{r}_4}=0$. 
Choosing the numerical values $m_H=450$\,GeV, $m_{\tilde{q}}=800$\,GeV,
$m_\chi=200$\,GeV, $\sqrt{s}=1700$\,GeV, fixing $s_{35}$ to $s_{35}=2(m_{\tilde{q}}^2+m_\chi^2)$
and combining with the phase space constraints of eq.~(\ref{psboundaries}) and the requirement 
$z\cdot {\cal S}\cdot z\leq 0$, 
we encounter the following discontinuities, shown in Fig.\,\ref{fig:Hbbsing}: 
\begin{itemize}
\item a normal threshold at $\sqrt{s_{45}}=(m_{\tilde{q}}+m_\chi)=1000$\,GeV, where 
$\lambda(s_{45},m_{\tilde{q}}^2,M_\chi^2)$ vanishes, corresponding to the production 
of a squark and a neutralino in the cut two-point diagram associated with a pinch of both $r_2$
and $r_4$ in the box diagram.
\item an anomalous threshold at $\sqrt{s_{45}}\simeq 1012.7$\,GeV, corresponding to \\
$\det {{\cal S}}^{(3)}_{\hat{r}_2}=0$.
\item an anomalous threshold at $\sqrt{s_{45}}\simeq 1038.1$\,GeV, corresponding to \\
$\det {{\cal S}}^{(3)}_{\hat{r}_4}=0$.
\item a leading Landau singularity at $\sqrt{s_{45}}\simeq 1078.4$\,GeV, corresponding to \\
$\det {{\cal S}}^{(4)}=0$.
\end{itemize}

\begin{figure}[htb]
\unitlength=1mm
\begin{center}
\begin{picture}(150,100)
\put(5,90){\includegraphics[width=8.5cm,angle=-90]{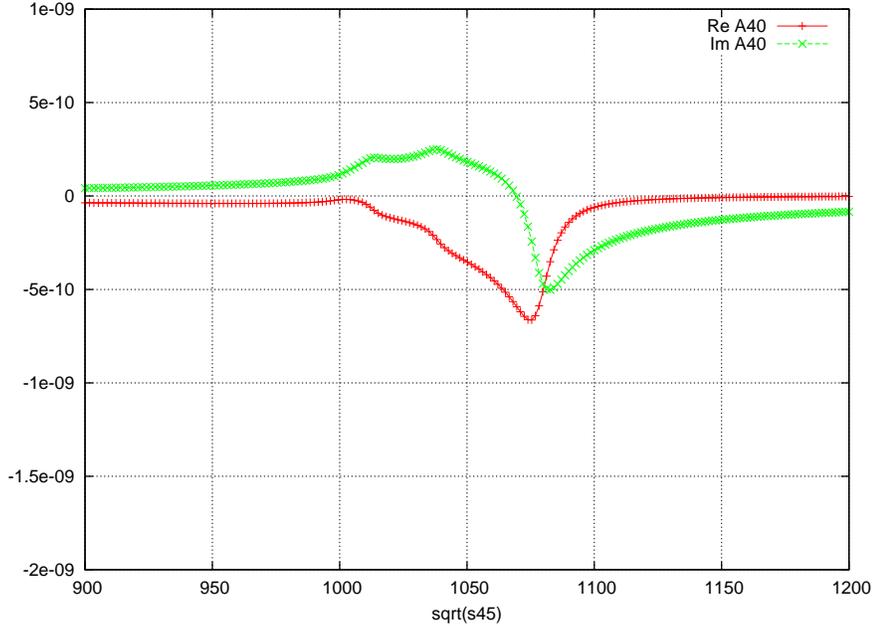}}
\end{picture}
\end{center}
\caption{Singularity structure of the scalar four-point function (real masses) contained in the diagram 
of Fig.\,\ref{fig:Hbb} for $m_H=450$\,GeV, 
$m_{\tilde{q}}^2\to m_{\tilde{q}}^2-i\,m_{\tilde{q}}\Gamma_{\tilde{q}}$, 
$m_\chi^2\to m_\chi^2-i\,m_\chi\Gamma_\chi$, 
$\Gamma_{\tilde{q}}=3.5\,{\rm{GeV}}, \Gamma_\chi=1.5$\,GeV }
\label{fig:HbbsingC}
\end{figure}

Introducing complex masses moves the poles due to propagators going on-shell away 
from the real axis and therefore regulates the leading Landau singularity, as can be seen 
from Fig.\,\ref{fig:HbbsingC}.

We should remark that Landau singularities due to massive (unstable) particles 
in the loop going on-shell
usually behave like $1/\sqrt{\det {\cal S}}$ as $\det {\cal S}\to 0$.
This is in contrast to cases with several massless particles, where 
$\det {\cal S}$ usually has several zero eigenvalues. 
An interesting case is the six-photon amplitude, where not only $\det {\cal S}$,  
but also its {\it derivative} with respect to the invariants involved vanishes at the 
singular point, meaning that the singularity is not integrable anymore. 
Introducing an imaginary part as advocated above would certainly not help in this case, 
but in the six-photon example we are saved by the gauge structure, which leads to numerators 
taming the singularity structure when the individual contributions are combined to  
physical helicity amplitudes. This has been worked out in detail in 
Refs.\,\cite{Nagy:2006xy,Bernicot:2007hs,Bern:2008ef}.

\subsection{Tensorial Reconstruction of the Integrand}
\label{ssec:tensrecth}

The library \texttt{golem95} in its original version\cite{Binoth:2008uq}, 
when used in amplitude calculations, relies
on some user generated code expressing a diagram\footnote{Actually, it is
not required to organize the calculation in terms of diagrams. The term
\emph{diagram} here refers to a set of terms of an amplitude sharing some
common loop propagators.} in terms of form factors
($A^{N,r}, B^{N,r}, C^{N,r}$)
as defined in Eq.~\eqref{eq:formfactordef}.
This requirement of the previous version of 
\texttt{golem95} restricted the applicability
of the library to algebraic methods for generating the amplitude.
In order to be able to use the library in the context of a 
 numerical reconstruction of the integrand, we included  new features 
 described in the following.
 
The general structure of a one-loop diagram can be written as
\begin{equation}
\mathcal{G}=\int\!\!\frac{\mathrm{d}^D q}{i\pi^{D/2}}\frac{\mathcal{N}(q)}%
{(q_1^2-m_1^2+i\delta)\cdots(q_N^2-m_N^2+i\delta)}
\end{equation}
where $q_a=q+r_a$. After decomposing the $D$-dimensional vector $q$ into its projection $\hat{q}$
onto the physical 4-dimensional Minkowski space and the radial component $\mu$
of the $(D-4)$-dimensional orthogonal space, such that $q^2=\hat{q}^2-\mu^2$, we can express
the numerator function $\mathcal{N}(q)$ in terms of the following tensor
structure,
\begin{equation}
\mathcal{N}(q)=\mathcal{N}(\hat{q}, \mu^2)=
\sum_{\alpha=0}^2\mu^{2\alpha}\left(C_{0,\alpha}+
\sum_{r=1}^{R-2\alpha} C_{r,\alpha}^{\mu_1\ldots\mu_r}
  \hat{q}_{\mu_1}\cdots\hat{q}_{\mu_r}\right)\;.
  \label{numstruc}
\end{equation}
The coefficients $C_{r,\alpha}^{\cdots}$ can be determined numerically
by the algorithm described in Ref.~\cite{Heinrich:2010ax}. The contraction
of the coefficients with the tensor integrals can be carried out in a
process independent and numerical way. Therefore this tensorial reconstruction
of the integrand provides a way of processing both diagrams from algebraic
constructions and purely numerical input, using the numerator function
$\mathcal{N}(\hat{q}, \mu^2)$ and the set of denominators as the only
common source of information. We have implemented the
reconstruction of the coefficients and their contraction with the
tensor integrals as a part of \texttt{golem95}. The new interface
is described in Section~\ref{ssec:tensrecint}.

The contraction with the tensor integrals also requires the implementation
of integrals with explicit $\mu^2$-dependence, which can be computed from
the poles of known form factors,
\begin{multline}
\int\!\!\frac{\mathrm{d}^Dq}{i\pi^{D/2}}
   \frac{\mu^2 q_{a_1}^{\mu_1}\cdots q_{a_{r-2}}^{\mu_{r-2}}}%
   {(q_1-m_1^2+i\delta)\cdots(q_N-m_N^2+i\delta)}=\\
	(4-D)\left(\sum_{j_1,\ldots,j_{r-2}\in S}
   \left[\Delta_{j_1\cdot}^{\cdot}\cdots\Delta_{j_{r-2}\cdot}^{\cdot}%
   \right]^{\{\mu_1\ldots\mu_{r-2}\}}_{\{a_1\ldots a_{r-2}\}}
   B^{N,r}_{j_1\ldots j_{r-2}}(S)\right.
\\
\left. + \sum_{j_1,\ldots,j_{r-4}\in S}
   \left[\hat{g}^{\cdot\cdot}%
   \Delta_{j_1\cdot}^{\cdot}\cdots\Delta_{j_{r-4}\cdot}^{\cdot}%
   \right]^{\{\mu_1\ldots\mu_{r-2}\}}_{\{a_1\ldots a_{r-2}\}}
   C^{N,r}_{j_1\ldots j_{r-4}}(S)
\right)+\mathcal{O}(D-4).
\end{multline}
In practice, the above equation simplifies greatly since the only
non-zero contributions come from the form factors $B^{2,2}, B^{3,2}, B^{3,3}$
and $C^{4,4}$. The only non-zero integral with $\mu^4$ in the numerator
is the box
\begin{multline}
\int\!\!\frac{\mathrm{d}^Dq}{i\pi^{D/2}}
   \frac{\mu^4}%
   {(q_1-m_1^2+i\delta)\cdots(q_4-m_4^2+i\delta)}=\\
  (D-4)(D-2)C^{4,4}(S)
 +\mathcal{O}(D-4).
\end{multline}


\section{Overview of the software structure}

The structure of the {\tt golem95} program is the following:
There are four  main directories:
\begin{enumerate}
\item {\bf src:} the source files of the program
\item {\bf demos:} some programs for demonstration
\item {\bf doc:} documentation which has been created with robodoc~\cite{robodoc} 
\item {\bf test:} supplements the 
demonstration programs, containing files to produce form factors with user-defined kinematics. 
      The user can specify the rank, numerator, numerical point 
      etc. via a steering file.
\end{enumerate}
The subdirectory structure is the same as described in \cite{Binoth:2008uq}.

\section{Description of the individual software components}

We focus here on the new features, for more details on the software 
components which are the same as in version 1.0, we refer to \cite{Binoth:2008uq}
and to the documentation contained in the program.

\subsection{Form factor evaluation}

A typical setup for the calculation of form factors is:

\begin{flushleft}
{\tt call initgolem95(dim)}\\
\vspace{2mm}
\dots fill matrix ${\cal S}$ \dots\\
\vspace{2mm}
{\tt call preparesmatrix()}\\
\vspace{2mm}
\dots evaluate form factors \dots\\
\vspace{2mm}
{\tt call exitgolem95()}
\end{flushleft}
\vspace{2mm}

The three subroutines act as interfaces to the library. They are used to initialize 
and manipulate objects and features of the program. They are implemented in the module 
\texttt{matrice\_s} and will be described in more detail in the following.
To pass the desired masses and momenta, the user assigns values to the 
object \texttt{s\_mat}, which are the entries of the kinematic matrix ${\cal S}$
given in eq.~(\ref{eqDEFS}).
The form factors are implemented in the respective modules in \texttt{src/form\_factor}. 
We will sketch the internal structure of the library and the distinction 
between real and complex masses in the following.

\subsubsection{The module \texttt{matrice\_s}}

This module is located in \texttt{src/kinematic/matrice\_s.f90}. 
It is used to reserve and free memory for the kinematic matrix $\cal{S}$ and related objects, 
as well as for the computation of the inverse matrices. The three macro functions described above are:

\begin{description}

\item[\texttt{subroutine initgolem95(dim, opt\_set)}] allocates memory for two arrays 
\texttt{s\_mat\_c} and \texttt{s\_mat\_r} of dimension \texttt{dim} and rank two. 
These arrays represent internally the complex valued matrix $\cal{S}$ and its real part. 
Additional memory is reserved for the inverse of $\cal{S}$ and its submatrices 
as well as for parameters appearing in the reductions of the form factors. 
A public pointer \texttt{s\_mat} is associated with \texttt{s\_mat\_c}. 
This object is used to assign the matrix $\cal{S}$, both in the real mass and complex mass case.
An additional argument \texttt{opt\_set} can be given, which is an integer 
array reflecting the numbering of the propagators. 
The default value is an integer vector of range 1 to \texttt{dim}. 
This subroutine also initializes the caching system.
\\

\item[\texttt{subroutine preparesmatrix()}] creates the derived type object \texttt{s\_mat\_p}. 
This consists of two pointers to \texttt{s\_mat\_c} and \texttt{s\_mat\_r} 
as well as two integers encoding information about the entries of complex masses and vanishing masses. 
We describe the usage of this object in more detail in  section \ref{istructure}.
The arrays used later for the inverse matrices and sub-matrices are calculated in this routine. 
If the user defines values for the  matrix $\cal{S}$ which are purely real, 
only the real arrays are needed and calculated.
\\

\item[\texttt{subroutine exitgolem95()}] deallocates all arrays, nullifies all pointers and clears the cache.

\end{description}

\vspace*{3mm}

Note that the subroutines {\tt initgolem95} and {\tt exitgolem95} also have been designed to 
make the setup more user-friendly, so the calls to the subroutines {\tt allocation\_s}, 
{\tt allocate\_cache}, {\tt  init\_invs},  {\tt deallocation\_s}, {\tt clear\_cache} and 
the definition of {\tt set\_ref} 
in version 1.0.0 are now obsolete.

\subsubsection{Internal structure}\label{istructure}
Internally, the library is subdivided into three layers, implementing the reduction from 
general form factors down to specific integrals. The global derived type object 
\texttt{s\_mat\_p} is passed from one layer to the next. The information encoded 
in the integer bits of \texttt{s\_mat\_p}, is used to direct the evaluation of the 
form factors  to the specific integrals needed.

Upon a form factor call, reductions are performed in the respective modules located 
in \texttt{src/form\_factor}. These reductions only need information from the inverse 
matrices and related objects created with \texttt{preparesmatrix()}.

From the form factor modules, generic $N$-point functions are called, passing the derived 
type object \texttt{s\_mat\_p}. At this level, further reductions might be performed. 
Up to this point, the only information needed from the kinematic matrix $\cal{S}$ is 
the position of the zero-mass entries, which is encoded in the integer \texttt{s\_mat\_p\%b\_zero}. 
In the current implementation, we choose massless six-dimensional boxes and massive four-dimensional 
boxes as our basis of the reduction. A comparison of \texttt{b\_zero} with the pinched propagators 
gives a fast case distinction.

Only in the last step, when the generic $N$-point functions call specific integrals, 
does the (sub-) matrix $\cal{S}$ need to be passed over. 
If the latter contains complex masses, which again can be quickly determined by comparing
 \texttt{s\_mat\_p\%b\_cmplx} with the set of remaining  propagators, 
 an implementation of the integral for complex masses is called with \texttt{s\_mat\_p\%pt\_cmplx}, 
 otherwise the version for real masses is called with \texttt{s\_mat\_p\%pt\_real}.
This structure of the library allows for an efficient reduction of the form factors to the basis integrals, 
without a full knowledge of the kinematic matrix $\cal{S}$. 

\subsection{Interface for tensorial reconstruction}
\label{ssec:tensrecint}
The tensorial reconstruction described in Section~\ref{ssec:tensrecth}
has been implemented in two modules, \texttt{tens\_rec} and
\texttt{tens\_comb}. In the simplest case the user only needs to access
the function \texttt{evaluate} from the module \texttt{tens\_comb}. For
more advanced applications of the interface the user will have to use
functions and data types of both modules.

\subsubsection{\bf The module \texttt{tens\_rec}} 
The module \texttt{tens\_rec} is located in
\texttt{src/interface/tens\_rec.f90} and contains all routines required
for the reconstruction of the tensor coefficients and for the evaluation
of reconstructed numerators.

\paragraph{Data types}
The module \texttt{tens\_rec} defines the data types
\begin{displaymath}
\mathtt{coeff\_type\_1,\ldots,coeff\_type\_6},
\end{displaymath} with components
\texttt{c}$k$ ($0\leq k\leq\min(4,R)$) to store the coefficients of
a numerator of maximum rank~$R$. In particular,
the component \texttt{c$k$} is a two dimensional complex array
which stores the coefficients of the monomials with
$k$ distinct components of $\hat{q}$. The first index of the array
labels the $\binom{4}{k}$ ways of choosing $k$ of the four
components of $\hat{q}$. The second index labels the
$\sum_{r=1}^R\binom{r-1}{ k-1}=\binom{R}{ k}$
monomials of the corresponding polynomial. A certain order of the
array entries should not be assumed by the user.

\paragraph{Functions and subroutines}
In all functions and subroutines which have a parameter \texttt{numeval}
it should point to a function with the following signature:
\begin{verbatim}
interface
   function numeval(Q, mu2)
      use precision_golem, only: ki
      real(ki), dimension(0:3), intent(in) :: Q
      real(ki), intent(in) :: mu2
      complex(ki) :: numeval
   end function
end interface
\end{verbatim}
For all momenta the order $\mathtt{k(0:3)}=(E,x,y,z)$ is assumed.

\begin{description}
\item[\texttt{subroutine reconstruct$r$(numeval,c0,c1,c2)}] for $r=1,\dots,6$
reconstructs the tensor coefficients of the numerator function given by
\texttt{numeval}. The parameters \texttt{c0}, \texttt{c1} and \texttt{c2}
are output parameters; \texttt{c1} and \texttt{c2} are optional and only
available in the subroutines where $r\geq2$. The argument \texttt{c0} is
of type \texttt{coeff\_type\_$r$} and contains the constant part (with
respect to $\mu^2$, i.e. $\alpha=0$ in eq.~(\ref{numstruc})) of the polynomial. 
The arguments \texttt{c1} and
\texttt{c2} are of type \texttt{coeff\_type\_$(r-2)$} and contain
the coefficients of the $\mu^2$ and $\mu^4$ part of the polynomial\footnote{%
There is no \texttt{coeff\_type\_0}: \texttt{complex(ki)} is used instead.}.
\item[\texttt{subroutine print\_coeffs(coeffs, unit)}] is overloaded and can
take any derived coefficient type as its first argument. The second argument
is optional and has the default value \texttt{unit=6} (standard output).
This routine pretty-prints the coefficients to the given file or~device.
\item[\texttt{pure function tenseval$r$(Q, coeffs, max\_k)}] evaluates
the polynomial given by the coefficients \texttt{coeffs} for the given
\emph{real} momentum \texttt{Q}.
The optional argument \texttt{max\_k} is only
used internally and should not be assigned by the~user.
\item[\texttt{pure function ctenseval$r$(Q, coeffs)}] evaluates
the polynomial given by the coefficients \texttt{coeffs} for the given
\emph{complex} momentum \texttt{Q}.
\end{description}

The subroutines \texttt{reconstruct$r$} and \texttt{ctenseval$r$}
can together serve as an implementation of a \emph{presampling} for programs of
unitarity based reductions
at the integrand level~\cite{Ossola:2007ax,Mastrolia:2010nb}
or as an implementation of
a rescue system when the unitarity based method fails due to vanishing
Gram determinants.
Both options are described in more detail in~\cite{Heinrich:2010ax}.

\subsubsection{\bf The module \texttt{tens\_comb}}
The module \texttt{tens\_comb} is located in
\texttt{src/interface/tens\_comb.f90} and contains all routines for the
contraction of reconstructed tensor coefficients with tensor integrals.
It also contains the convenience function \texttt{evaluate} which combines
the steps required for reconstruction and contraction.

\paragraph{Functions and subroutines}
\begin{description}
\item[\texttt{function evaluate(numeval, momenta, set, rank)}] combines
tensorial reconstruction and the contraction of the coefficients with the
tensor integrals. The argument \texttt{momenta} is of
\texttt{dimension(:,0:3)} and contains the momenta $r_1,\ldots, r_N$.
The  argument {\tt set} can either be an integer array or an integer number
(generated by the function \texttt{packb}) and denotes the set of pinched
propagators; a value of zero can be used if no propagators are pinched.
It should be noted that the argument \texttt{momenta} also includes the
momenta belonging to pinched propagators. The last argument, \texttt{rank}, is optional.
It specifies the rank of the numerator; if omitted it is assumed that
$\mathtt{rank}=N$.
\item[\texttt{function contract$N$\_$r$(coeffs, momenta, set)}]
contracts
the coefficients\linebreak[4] \texttt{coeffs} of type \texttt{coeff\_type\_$r$} with
the $N$-point rank-$r$ tensor integral. The arguments \texttt{momenta}
and \texttt{set} are defined as in \texttt{evaluate} with the restriction
that \texttt{set} has to be an integer number.
\item[\texttt{function contract$N$\_$r$s1(coeffs, momenta, set)}]
contracts
the coefficients \texttt{coeffs} of type \texttt{coeff\_type\_$(r-2)$} with
the corresponding $N$-point tensor integral containing $\mu^2$ in the numerator.
The arguments \texttt{momenta}
and \texttt{set} are defined as in \texttt{evaluate} with the restriction
that \texttt{set} has to be an integer number.
\item[\texttt{function contract$N$\_$r$s2(coeffs, momenta, set)}]
contracts
the coefficients \texttt{coeffs} of type \texttt{coeff\_type\_$(r-2)$} with
the corresponding $N$-point tensor integral containing $\mu^4$ in the numerator.
The arguments \texttt{momenta}
and \texttt{set} are defined as in \texttt{evaluate} with the restriction
that \texttt{set} has to be an integer number.
\end{description}

The function \texttt{evaluate} determines $N$ from the size of
\texttt{momenta} and the number of pinches in \texttt{set}. It
then combines the according calls to \texttt{reconstruct$r$} and
\texttt{contract$N$\_$r$},
\texttt{contract$N$\_$r$s1} and
\texttt{contract$N$\_$r$s2}. The functions \texttt{evaluate} and
\texttt{contract\ldots} require that the matrix \texttt{s\_mat} and
the cache are set up properly by the sequence which one would have to
call when evaluating form factors.

\section{Installation instructions}

The program can be downloaded as an archive golem95-1.2.0.tar.gz from the following URL:
{\tt http://projects.hepforge.org/golem/95/}.
The installation instructions given below also can be found in the
{\tt Readme} file which comes with the code.
Information and updates of the program can also be found at 
{\tt http://projects.hepforge.org/golem/trac/wiki/golem95C}.

The installation setup is based on  autotools~\cite{Autotools}. 
To install the {\tt golem95} library, type the following commands:\\
{\tt ./configure [--prefix=mypath] [--precision=quadruple] [FC=compiler] [F77=fortran77compiler]}\\
{\tt make}\\
{\tt make install}

The \texttt{--prefix} option denotes the installation prefix, under which
the directories \texttt{lib/} and \texttt{include/} are generated. If no
option is given, on a Linux system the configure script would choose
\texttt{--prefix=/usr/local}. The argument \texttt{--precision} selects
if double or quadruple precision should be used in the library; it should
be noted that quadruple precision is not supported by all Fortran compilers
and therefore \texttt{--precision=double} is the default value.
If the variable \texttt{FC} is not set the first fortran compiler which is
automatically detected will be used. Another variable commonly used is
\texttt{FCFLAGS} which allows one to pass compiler flags to the Fortran compiler.
The fortran77 compiler is only needed to run the demo file {\tt demos/demo\_LT.f}.

In addition, it is possible to call the finite scalar box and triangle integrals 
with internal masses from LoopTools. The setup to do so is automated, all the user has to do
for this option is (a) install LoopTools, and (b) use the option 
{\tt [--with-looptools=your\_path\_to\_libooptools.a]} for the configure script, i.e. 
type the following commands:\\
{\tt ./configure [--prefix=mypath] [--with-looptools=your\_path\_to\_libooptools.a] 
[--precision=quadruple] [FC=compiler] [F77= fortran77compiler]}\\
{\tt make}\\
{\tt make install}.

\section{Examples}

Examples can be found in the subdirectory {\tt demos}. 
The program \texttt{demos/}\hspace{0cm}\texttt{demo\_cmplx\_masses.f90} shows
how the library is used to  evaluate form factors.
For a given matrix $\cal{S}$ with real or complex masses, 
six-point form factors from rank 0 to rank 6, as well as one- to five-point form factors are evaluated.
The results are written to the files \texttt{test\_ff(6)\_c/r.txt}.

The program \texttt{demos/demo\_tens\_rec.f90} together with the module
\texttt{demos/}\hspace{0cm}\texttt{demo\_tens\_mod.f90} demonstrates the use of the function
\texttt{evaluate}.
In this example,  a toy amplitude is given by a six-point integral
 where the numerator consists of three propagators which are also present 
 in the denominator.
 The scalar three-point function resulting from the direct cancellation of 
 three propagators is compared to the results obtained by expanding
 the numerator into contracted loop and external momenta, leading to 
 rank 6 hexagons if no propagators are cancelled, 
 rank 4 pentagons if one propagator is cancelled,
 rank 2 boxes if two propagators are cancelled.

The program used to scan the Landau singularity as in Fig.\,\ref{fig:HbbsingC} can be found in 
 \texttt{demos/SusyLandau.f90}.

\section{Conclusions}
We have presented a program for the numerical evaluation of scalar 
integrals and tensor form factors 
entering the calculation of one-loop amplitudes, 
which is able to provide results for real as well as 
complex masses in the loop integrals.
The program is an extension of on an earlier version of the 
golem95 library, but now also can be used in the context of a
unitarity-inspired numerical reconstruction of the integrand at the tensorial level.
Improvements in the caching system and in the user interface also have been made.
The program, available at 
{\tt http://projects.hepforge.org/golem/95/}, 
provides a complete library of scalar and tensor integrals 
up to rank six 6-point functions, including the option of complex masses.

\section{Acknowledgements}
We would like to thank Thomas Hahn for reading the manuscript and useful comments.
This research was supported by the UK Science and Technology Facilities Council 
(STFC) and the Scottish Universities Physics Alliance (SUPA).
T.R. has been supported by the Foundation FOM, project
FORM 07PR2556.


\begin{thebibliography}{99}
\bibitem{vanHameren:2009dr}
A.~van Hameren, C.~G. Papadopoulos, and R.~Pittau, {\it {Automated one-loop
  calculations: a proof of concept}},  {\em JHEP} {\bf 09} (2009) 106,
  [{{\tt arXiv:0903.4665}}].

\bibitem{vanHameren:2010cp}
A.~van Hameren, {\it {OneLOop: for the evaluation of one-loop scalar
  functions}}, {{\tt
  arXiv:1007.4716}}.

\bibitem{Hahn:1998yk}
T.~Hahn and M.~Perez-Victoria, {\it {Automatized one-loop calculations in four
  and D dimensions}},  {\em Comput. Phys. Commun.} {\bf 118} (1999) 153--165,
  [{{\tt hep-ph/9807565}}].

\bibitem{Hahn:2010zi}
T.~Hahn, {\it {Feynman Diagram Calculations with FeynArts, FormCalc, and
  LoopTools}},  {{\tt
  arXiv:1006.2231}}.

\bibitem{Bern:2008ef}
{\bf NLO Multileg Working Group} Collaboration, Z.~Bern {\em et.~al.}, {\it
  {The NLO multileg working group: summary report}},
  {{\tt arXiv:0803.0494}}.

\bibitem{Binoth:2010ra}
{\bf SM and NLO Multileg Working Group} Collaboration, J.~R. Andersen {\em
  et.~al.}, {\it {The SM and NLO multileg working group: Summary report}},
  {{\tt arXiv:1003.1241}}.

\bibitem{Ossola:2007ax}
G.~Ossola, C.~G. Papadopoulos, and R.~Pittau, {\it {CutTools: a program
  implementing the OPP reduction method to compute one-loop amplitudes}},  {\em
  JHEP} {\bf 03} (2008) 042, [{{\tt arXiv:0711.3596}}].

\bibitem{Mastrolia:2010nb}
P.~Mastrolia, G.~Ossola, T.~Reiter, and F.~Tramontano, {\it {Scattering
  AMplitudes from Unitarity-based Reduction Algorithm at the Integrand-level}},
   {\em JHEP} {\bf 08} (2010) 080,
  [{{\tt arXiv:1006.0710}}].

\bibitem{Heinrich:2010ax}
G.~Heinrich, G.~Ossola, T.~Reiter, and F.~Tramontano, {\it {Tensorial
  Reconstruction at the Integrand Level}},  {\em JHEP} {\bf 10} (2010) 105,
  [{{\tt arXiv:1008.2441}}].

\bibitem{Hahn:2000kx}
T.~Hahn, {\it {Generating Feynman diagrams and amplitudes with FeynArts 3}},
  {\em Comput. Phys. Commun.} {\bf 140} (2001) 418--431,
  [{{\tt hep-ph/0012260}}].

\bibitem{Hahn:2006qw}
T.~Hahn and M.~Rauch, {\it {News from FormCalc and LoopTools}},  {\em Nucl.
  Phys. Proc. Suppl.} {\bf 157} (2006) 236--240,
  [{{\tt hep-ph/0601248}}].

\bibitem{Campbell:2002tg}
J.~Campbell and R.~K. Ellis, {\it {Next-to-leading order corrections to W +
  2jet and Z + 2jet production at hadron colliders}},  {\em Phys. Rev.} {\bf
  D65} (2002) 113007, [{{\tt hep-ph/0202176}}].

\bibitem{Arnold:2008rz}
K.~Arnold {\em et.~al.}, {\it {VBFNLO: A parton level Monte Carlo for processes
  with electroweak bosons}},  {\em Comput. Phys. Commun.} {\bf 180} (2009)
  1661--1670, [{{\tt arXiv:0811.4559}}].

\bibitem{vanOldenborgh:1989wn}
G.~J. van Oldenborgh and J.~A.~M. Vermaseren, {\it {New Algorithms for One Loop
  Integrals}},  {\em Z. Phys.} {\bf C46} (1990) 425--438.

\bibitem{vanOldenborgh:1990yc}
G.~J. van Oldenborgh, {\it {FF: A Package to evaluate one loop Feynman
  diagrams}},  {\em Comput. Phys. Commun.} {\bf 66} (1991) 1--15.

\bibitem{Ellis:2007qk}
R.~K. Ellis and G.~Zanderighi, {\it {Scalar one-loop integrals for QCD}},  {\em
  JHEP} {\bf 02} (2008) 002, [{{\tt arXiv:0712.1851}}].

\bibitem{Binoth:2008uq}
T.~Binoth, J.~P. Guillet, G.~Heinrich, E.~Pilon, and T.~Reiter, {\it {Golem95:
  a numerical program to calculate one-loop tensor integrals with up to six
  external legs}},  {\em Comput. Phys. Commun.} {\bf 180} (2009) 2317--2330,
  [{{\tt arXiv:0810.0992}}].

\bibitem{Diakonidis:2010rs}
T.~Diakonidis, J.~Fleischer, T.~Riemann, and B.~Tausk, {\it {A recursive
  approach to the reduction of tensor Feynman integrals}},  {\em PoS} {\bf
  RADCOR2009} (2010) 033, [{{\tt arXiv:1002.0529}}].

\bibitem{Nhung:2009pm}
D.~T. Nhung and L.~D. Ninh, {\it {D0C : A code to calculate scalar one-loop
  four-point integrals with complex masses}},  {\em Comput. Phys. Commun.} {\bf
  180} (2009) 2258--2267, [{{\tt arXiv:0902.0325}}].

\bibitem{Denner:2010tr}
A.~Denner and S.~Dittmaier, {\it {Scalar one-loop 4-point integrals}},
  {{\tt arXiv:1005.2076}}.

\bibitem{tHooft:1978xw}
G.~'t~Hooft and M.~J.~G. Veltman, {\it {Scalar One Loop Integrals}},  {\em
  Nucl. Phys.} {\bf B153} (1979) 365--401.

\bibitem{Fabricius:1979tb}
K.~Fabricius and I.~Schmitt, {\it {Calculation Of Dimensionally Regularized Box
  Graphs In The Zero Mass Case}},  {\em Z. Phys.} {\bf C3} (1979) 51--53.

\bibitem{Beenakker:1988jr}
W.~Beenakker and A.~Denner, {\it {Infrared Divergent Scalar Box Integrals With
  Applications In The Electroweak Standard Model}},  {\em Nucl. Phys.} {\bf
  B338} (1990) 349--370.

\bibitem{Denner:1991qq}
A.~Denner, U.~Nierste, and R.~Scharf, {\it {A Compact expression for the scalar
  one loop four point function}},  {\em Nucl. Phys.} {\bf B367} (1991)
  637--656.

\bibitem{Bern:1992em}
Z.~Bern, L.~J. Dixon, and D.~A. Kosower, {\it {Dimensionally regulated one loop
  integrals}},  {\em Phys. Lett.} {\bf B302} (1993) 299--308,
  [{{\tt hep-ph/9212308}}].

\bibitem{Bern:1993kr}
Z.~Bern, L.~J. Dixon, and D.~A. Kosower, {\it {Dimensionally regulated pentagon
  integrals}},  {\em Nucl. Phys.} {\bf B412} (1994) 751--816,
  [{{\tt hep-ph/9306240}}].

\bibitem{Denner:1999gp}
A.~Denner, S.~Dittmaier, M.~Roth, and D.~Wackeroth, {\it {Predictions for all
  processes $e^+ e^- \to$ 4 fermions + gamma}},  {\em Nucl. Phys.} {\bf B560}
  (1999) 33--65, [{{\tt hep-ph/9904472}}].

\bibitem{Denner:2005fg}
A.~Denner, S.~Dittmaier, M.~Roth, and L.~H. Wieders, {\it {Electroweak
  corrections to charged-current $e^+ e^- \to$ 4 fermion processes: Technical
  details and further results}},  {\em Nucl. Phys.} {\bf B724} (2005) 247--294,
  [{{\tt hep-ph/0505042}}].

\bibitem{Actis:2008uh}
S.~Actis, G.~Passarino, C.~Sturm, and S.~Uccirati, {\it {Two-Loop Threshold
  Singularities, Unstable Particles and Complex Masses}},  {\em Phys. Lett.}
  {\bf B669} (2008) 62--68, [{{\tt arXiv:0809.1302}}].

\bibitem{Passarino:2010qk}
G.~Passarino, C.~Sturm, and S.~Uccirati, {\it {Higgs Pseudo-Observables, Second
  Riemann Sheet and All That}},  {\em Nucl. Phys.} {\bf B834} (2010) 77--115,
  [{{\tt arXiv:1001.3360}}].

\bibitem{Binoth:2005ff}
T.~Binoth, J.~P. Guillet, G.~Heinrich, E.~Pilon, and C.~Schubert, {\it {An
  algebraic/numerical formalism for one-loop multi-leg amplitudes}},  {\em
  JHEP} {\bf 10} (2005) 015,
  [{{\tt hep-ph/0504267}}].

\bibitem{Binoth:1999sp}
T.~Binoth, J.~P. Guillet, and G.~Heinrich, {\it {Reduction formalism for
  dimensionally regulated one-loop N-point integrals}},  {\em Nucl. Phys.} {\bf
  B572} (2000) 361--386, [{{\tt hep-ph/9911342}}].

\bibitem{Duplancic:2003tv}
G.~Duplancic and B.~Nizic, {\it {Reduction method for dimensionally regulated
  one-loop N- point Feynman integrals}},  {\em Eur. Phys. J.} {\bf C35} (2004)
  105--118, [{{\tt hep-ph/0303184}}].

\bibitem{Giele:2004iy}
W.~T. Giele and E.~W.~N. Glover, {\it {A calculational formalism for one-loop
  integrals}},  {\em JHEP} {\bf 04} (2004) 029,
  [{{\tt hep-ph/0402152}}].

\bibitem{delAguila:2004nf}
F.~del Aguila and R.~Pittau, {\it {Recursive numerical calculus of one-loop
  tensor integrals}},  {\em JHEP} {\bf 07} (2004) 017,
  [{{\tt hep-ph/0404120}}].

\bibitem{vanHameren:2005ed}
A.~van Hameren, J.~Vollinga, and S.~Weinzierl, {\it {Automated computation of
  one-loop integrals in massless theories}},  {\em Eur. Phys. J.} {\bf C41}
  (2005) 361--375, [{{\tt hep-ph/0502165}}].

\bibitem{Denner:2005nn}
A.~Denner and S.~Dittmaier, {\it {Reduction schemes for one-loop tensor
  integrals}},  {\em Nucl. Phys.} {\bf B734} (2006) 62--115,
  [{{\tt hep-ph/0509141}}].

\bibitem{Fleischer:2010sq}
J.~Fleischer and T.~Riemann, {\it {A complete algebraic reduction of one-loop
  tensor Feynman integrals}},  {{\tt arXiv:1009.4436}}.

\bibitem{Davydychev:1991va}
A.~I. Davydychev, {\it {A Simple formula for reducing Feynman diagrams to
  scalar integrals}},  {\em Phys. Lett.} {\bf B263} (1991) 107--111.

\bibitem{BjorkenDrell}
J.~Bjorken and S.~Drell, {\em {Relativistic Quantum Field Theory}}.
\newblock McGraw-Hill, New York, 1965.

\bibitem{Goria:2008ny}
S.~Goria and G.~Passarino, {\it {Anomalous Threshold as the Pivot of Feynman
  Amplitudes}},  {\em Nucl. Phys. Proc. Suppl.} {\bf 183} (2008) 320--325,
  [{{\tt arXiv:0807.0698}}].

\bibitem{Denner:1996ug}
A.~Denner, S.~Dittmaier, and T.~Hahn, {\it {Radiative corrections to Z Z $\to$
  Z Z in the electroweak standard model}},  {\em Phys. Rev.} {\bf D56} (1997)
  117--134, [{{\tt hep-ph/9612390}}].

\bibitem{Boudjema:2008zn}
F.~Boudjema and L.~D. Ninh, {\it {b anti-b Higgs production at the LHC: Yukawa
  corrections and the leading Landau singularity}},  {\em Phys. Rev.} {\bf D78}
  (2008) 093005, [{{\tt arXiv:0806.1498}}].

\bibitem{Nagy:2006xy}
Z.~Nagy and D.~E. Soper, {\it {Numerical integration of one-loop Feynman
  diagrams for N- photon amplitudes}},  {\em Phys. Rev.} {\bf D74} (2006)
  093006, [{{\tt hep-ph/0610028}}].

\bibitem{Bernicot:2007hs}
C.~Bernicot and J.~P. Guillet, {\it {Six-Photon Amplitudes in Scalar QED}},
  {\em JHEP} {\bf 01} (2008) 059,
  [{{\tt arXiv:0711.4713}}].

\bibitem{robodoc}
http://www.xs4all.nl/$\sim$rfsber/robo/robodoc.html.

\bibitem{Autotools}
G.~V. Vaughan, B.~Elliston, T.~Tromey, and I.~L. Taylor, {\em GNU Autoconf,
  Automake, and Libtool: Expert insight into porting software and building
  large projects using GNU Autotools}.
\newblock New Riders, Indianapolis, 2000.

\end{thebibliography}
\end{document}